\documentclass{webofc}
\usepackage[varg]{txfonts}   
\begin{document}
\title{Confirmation of NIKA2 investigation of the Sunyaev-Zel'dovich effect by using synthetic clusters of galaxies}
%
%

\author{\firstname{Marco} \lastname{De Petris}\inst{1}\fnsep\thanks{\email{marco.depetris@roma1.infn.it}} \and 
	\firstname{Florian} \lastname{Ruppin}\inst{2} 
	\and
	\firstname{Federico} \lastname{Sembolini}\inst{1,3,4}
	\and
	\firstname{Rem\'i} \lastname{Adam}\inst{5}
	\and
	\firstname{Anna Silvia} \lastname{Baldi}\inst{1,6}
	\and
	\firstname{Giammarco} \lastname{Cialone}\inst{1}
	\and
	\firstname{Barbara} \lastname{Comis}\inst{2}
	\and
	\firstname{Federico} \lastname{De Luca}\inst{1}
	\and
	\firstname{Giulia} \lastname{Gianfagna}\inst{1}
	\and
	\firstname{Florian} \lastname{K\'eruzor\'e}\inst{2}
	\and
	\firstname{Juan} \lastname{Mac\'ias-P\'erez}\inst{2}
	\and
	\firstname{Fr\'ed\'eric} \lastname{Mayet}\inst{2}
	\and
	\firstname{Laurence} \lastname{Perotto}\inst{2}
	\and
	\firstname{Gustavo} \lastname{Yepes}\inst{3,4}
}

\institute{Dipartimento di Fisica, Sapienza Universit\`a di Roma, Piazzale Aldo Moro 5, I-00185 Roma, Italy 
\and
Univ. Grenoble Alpes, CNRS, Grenoble INP, LPSC-IN2P3, 53, avenue des Martyrs, 38000 Grenoble, France
\and
          Departamento de F\'isica Te\'orica,M\'odulo 8, Facultad de Ciencias, Universidad Aut\'onoma de Madrid, E-28049 Cantoblanco, Madrid, Spain
\and
Centro de Investigaci\'on Avanzada en F\'isica Fundamental (CIAFF), Universidad Aut\'onoma de Madrid, E-28049 Madrid, Spain
\and
Laboratoire Leprince-Ringuet, \'Ecole Polytechnique, CNRS/IN2P3, 91128 Palaiseau, France
\and
Dipartimento di Fisica, Universit\`a di Roma “Tor Vergata”, Via della Ricerca Scientifica, I-00133 Roma, Italy
}

\abstract{%
The NIKA2 Sunyaev-Zel'dovich Large Program (SZLP) is focused on mapping the thermal SZ signal of a representative sample of selected \textsl{Planck} and ACT clusters spanning the redshift range 0.5<$z$<0.9. 
Hydrodynamical N-body simulations prove to be a powerful tool to endorse NIKA2 capabilities for estimating the impact of IntraCluster Medium (ICM) disturbances when recovering the pressure radial profiles. For this goal we employ a subsample of objects, carefully extracted from the catalog {\it Marenostrum MUltidark SImulations of galaxy Clusters} (MUSIC), spanning equivalent redshift and mass ranges as the LPSZ. The joint analysis of real observations of the tSZ with NIKA2 and \textsl{Planck} enables to validate the NIKA2 pipeline and to estimate the ICM pressure profiles. Moreover, the possibility to identify {\it a priori} the dynamical state of the selected synthetic clusters allows us to verify the impact on the recovered ICM profile shapes and their scatters. Morphological analysis of maps of the Compton parameter seems to be a way to observationally segregate the sample based on the dynamical state in relaxed and disturbed synthetic clusters.
}
\maketitle
\section{Introduction}
\label{intro}
Clusters of galaxies are a powerful target to provide useful cosmological information. The abundance of these objects in the Universe, as function of total mass and redshift, is related to the mean matter density, $\Omega_m$, and the amplitude of matter perturbations at a scale of 8$h^{-1}$Mpc, quantified with $\sigma_8$ ($e.g.$ \cite{Planck16}). The Sunyaev-Zel'dovich (SZ) effect, mainly the thermal component (tSZ) $i.e.$ the inverse Compton scattering of CMB photons with hot electron gas \cite{SZ72}, is a suitable probe to map the ICM pressure distribution in clusters even at high redshifts. Combining this information with X-ray data it is possible to infer cluster masses under the assumption of hydrostatic equilibrium. Unfortunately, besides neglecting non-thermal pressure contributions, such as bulk motions and ICM turbulence, these results can also be biased due to inaccurate knowledge of ICM pressure profiles in shape, scatter and redshift evolution.
In the case of large surveys, used to infer cosmological information, the overall cluster mass is directly inferred from the integrated tSZ flux over a solid angle being proportional to the thermal energy content of the galaxy clusters (see \cite{Arnaud10} and \cite{Planck14}).

\section{NIKA2 tSZ Large Program}
\label{tSZLP}
The NIKA2 tSZ Large Program (SZLP) consists of 300 hours of Guaranteed Time at the 30m-IRAM telescope to map the tSZ signal of a representative sample of 45 galaxy clusters (selected from \textsl{Planck} and ACT catalogues) in the redshift range 0.5<$z$<0.9 at high angular resolution (<20”). 19 clusters have been already observed by NIKA2 \cite{Adam18}. X-ray and optical follow-ups, with XMM-$Newton$ satellite and Canary Islands optical observatories, also provide useful additional information; see \cite{Mayet19} for all the details of the program.
Among the several objectives of the NIKA2 SZLP there is the characterization of the mean ICM pressure profile properties and of the systematic uncertainties associated with departures from a simple model of high redshift objects in hydrostatic equilibrium allowing unbiased cosmological results \cite{Ruppin19}. Hydrodynamical simulations seem to be a valuable test-bed for this goal.

\section{MUSIC dataset}
\label{MUSIC2}
Numerical hydrodynamical simulations are widely applied in Cosmology because they take into account nonlinearities in the large structure formation processes, allowing to estimate the expected shape of the mass function given the abundance of halos with different mass and along the redshift ($e.g.$ \cite{Tinker08}). Moreover, the self-similar scaling laws linking the tSZ observable with cluster mass are based on a pure gravitational collapse scenario of cluster formation whose deviations can be investigated through simulations (see \cite{Kra12} for a review).

\begin{figure*}[t]
\centering
\includegraphics[width=1\textwidth]{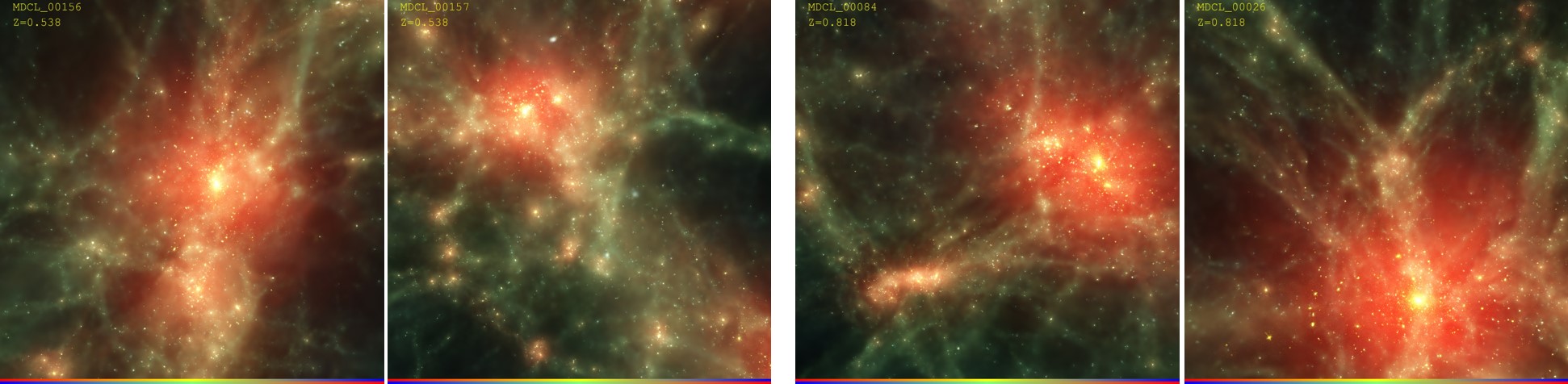}
\caption{Gas distribution in MUSIC-2 clusters simulated with radiative physics:  clu\#156 and 157 at $z$=0.54 ({\it Left}) and clu\#84 and 26 at $z$=0.82 ({\it Right}).}
\label{fig-0}
\end{figure*}

The synthetic clusters under study were extracted from the cosmological N-body dark matter-only simulation {\it MultiDark} \cite{Prada12}, a cube of 1$h^{-1}$Gpc aside. The {\it Marenostrum MUltidark SImulations of galaxy Clusters} (MUSIC-2) dataset is composed of the 283 most massive clusters ($M_{vir}$>10$^{15}h^{-1}$M$_\odot$ at $z$=0) resimulated with higher mass resolution and adding gas particles. Physical processes, such as gas cooling, stars formation and supernovae feedback, have been included. Each cluster is inside a sphere with a radius of 6$h^{-1}$Mpc at $z$=0 with mass resolution equal to $m_{DM}$=9.0x10$^{8}h^{-1}$M$_\odot$ and $m_{g}$=1.9x10$^{8}h^{-1}$M$_\odot$ for dark matter and gas mass particles, respectively \cite{Sembo13}. The assumed cosmological model is constrained by WMAP7 \cite{Komo11}: $\Omega_m$=0.27, $\Omega_b$=0.0469, $\Omega_{\Lambda}$=0.73, $h$=0.7, $\sigma_8$=0.82, and $n$=0.95. Among the available snapshots at different redshifts starting from $z$=9, we selected 32 clusters to populate the {\sl twin sample} of the NIKA2 SZLP one, covering redshift-mass plane at two redshifts, $z$=0.54 and $z$=0.82. Four of such clusters are shown in Fig.~\ref{fig-0}, while the whole sample distribution is represented in the left panel of Fig.~\ref{fig-1}. Being a volume-limited sample of synthetic clusters, this {\sl twin sample} features a limited number of high-mass objects at high redshift. For each cluster we have a map of the Compton parameter and radial profiles of ICM and dark matter. An advantage of the MUSIC-simulation approach is the availability of all the information to segregate the dynamical state of each object, see Sect.~\ref{morpho}.

\section{y-maps morphology and dynamical state}
\label{morpho}

\begin{figure*}[h]
\centering
\includegraphics[scale=0.45]{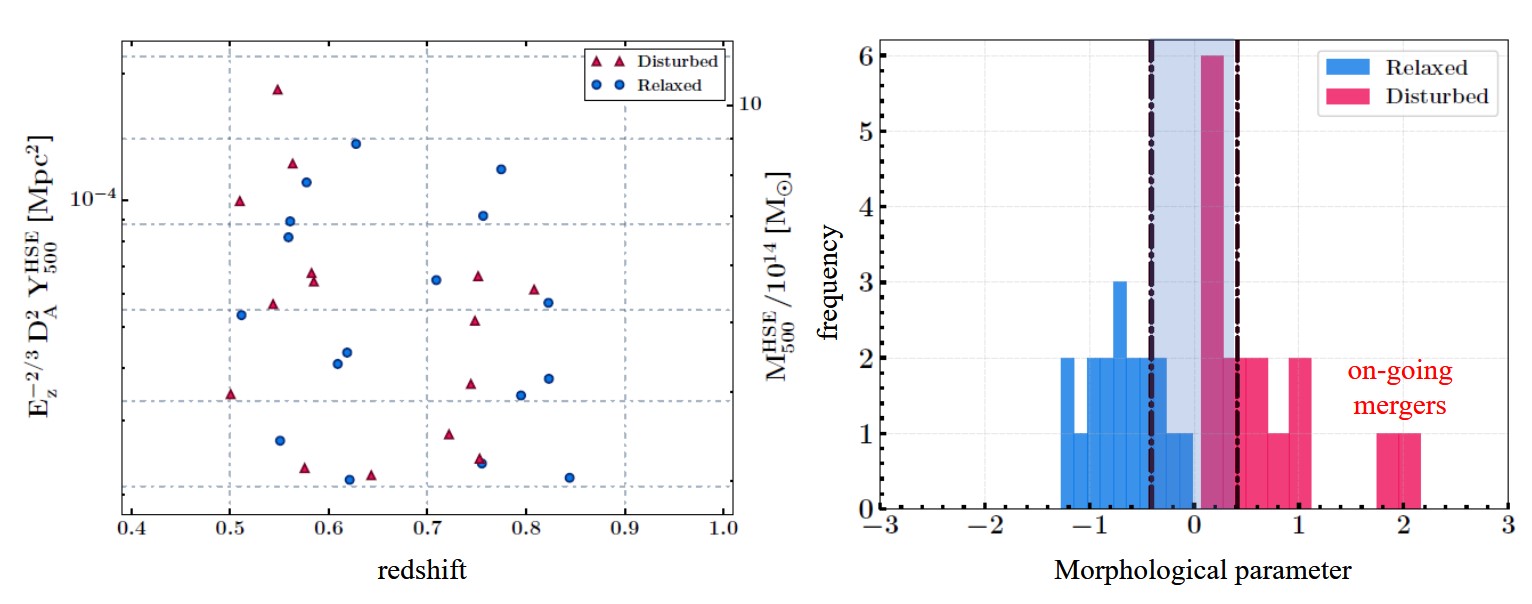}
\caption{{\it Left}: MUSIC clusters selected in the mass-redshift plane similarly to the targets of the NIKA2 SZLP. Morphologically relaxed and disturbed clusters are indicated by blue dots and red triangles respectively. The redshifts are randomly distributed around the two values for display reasons. {\it Right}: Distributions of the morphological indicators: in blue the relaxed and in red the disturbed sub-samples. The vertical band shows the region of hybrid objects \cite{Florian19}.}
\label{fig-1}
\end{figure*}

The maps of the Compton parameter, $y$-maps, are generated by integrating the electron pressure along a line of sight inside a cylinder of 6 $R_{vir}$ in height and in diameter \cite{Sembo13}. Each map has a size of 10 Mpc with pixels of 10 kpc. At the considered redshifts this results in a region of 25'x25' (larger than NIKA2 FOV of 6.5') with a resolution around 2" (lower than NIKA2 150GHz resolution of 18").
Among the selected 32 clusters, we identified a homogeneous dynamical state population by applying a morphological analysis of the $y$-maps, already tuned on MUSIC clusters as discussed in \cite{Cia18}.
The SZ morphology is studied by means of the morphological parameter, $M$, derived from the combination of 6 different 2D estimators (asymmetry parameter, light concentration, third-order power ratio, centroid shift, strip parameter and gaussian fit parameter), a few of them widely applied on X-ray maps. Each parameter contributes to $M$ with a different weight, carefully inferred by a Kolmogorov–Smirnov test on the distributions of the populations of relaxed and disturbed objects, as identified by two 3D indicators of the cluster dynamical state. They are: 1) $M_{sub}/M_{vir}$, the ratio between the mass of the largest sub-structure in the cluster and the virial mass and 2) $\Delta_{r}$, the offset between the positions of the peak of the density distribution and the centre of mass of the cluster, normalized to the virial radius. In both cases, a threshold at 0.1 is used to discriminate relaxed (<0.1) from disturbed objects (>0.1). We stress that the morphological parameter is prone to projection effects and our definition of ($M_{sub}/M_{vir}$) is biased in selecting merging processes. As a general criterion, we find that relaxed clusters show $M$<0.4, hybrid clusters have -0.4<$M$<0.4 and disturbed clusters $M$>0.4. The distribution of the dynamical state of the {\sl twin sample}, quantified with $M$, is shown in the right panel of Fig.~\ref{fig-1}.
Clusters dynamical state could also be inferred observationally, by estimating the offsets between 5 peculiar positions: Bright Central Galaxy, centroids and peaks of tSZ and X-ray maps \cite{DeLuca19}.

\section{NIKA2 and \textsl{Planck} simulated observations}
\label{NIKA2sims}
\begin{figure*}[t]
\centering
\includegraphics[scale=0.45]{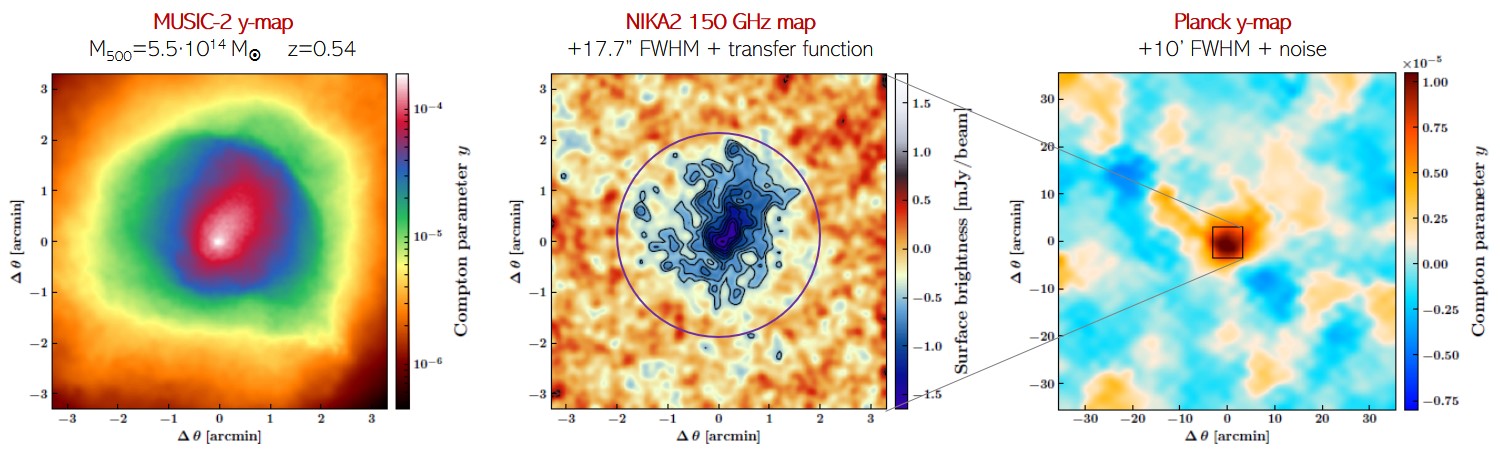}
\caption{{\it Left}: MUSIC Compton parameter map of a selected disturbed cluster at $z$=0.54. {\it Middle}: simulated NIKA2 tSZ surface brightness map at 150 GHz. The circle with r$\sim$2' is the well-mapped NIKA2 region. {\it Right}: simulated \textsl{Planck} Compton parameter map. The field of view of 6.5' considered for the left and middle panels is shown as a black square at the center of this map \cite{Florian19}.}
\label{fig-2}
\end{figure*}

The description of the study of MUSIC {\sl twin sample} by NIKA2 and \textsl{Planck} observations is fully detailed in \cite{Florian19}. We recall here the main steps. NIKA2 and \textsl{Planck} realistic tSZ observations towards each of the {\sl twin sample} clusters have been generated to be jointly analyzed to infer ICM pressure profiles. The Compton parameter map of an example cluster is shown in Fig.~\ref{fig-2} together with mock maps of NIKA2 (tSZ brightness at 150 GHz convolved by 17.7" FWHM Gaussian beam and the NIKA2 transfer function) and \textsl{Planck} ($y$-map convolved by 10' FWHM Gaussian beam). Residual noise is added by starting from the noise power spectra of each instrument. It is clear that NIKA2 accurately maps tSZ signal up to 2 arcmin from the cluster peak while \textsl{Planck} detection is useful to put some constraints on the total integrated Compton parameter, $Y$, over a larger aperture, 5$R_{500}$. 
CMB signal is negligible at NIKA2 angular scales while the Cosmic Infrared Background signal is lower than instrumental and atmospheric noises. Furthermore, no contaminants are included in order to study only the impact of ICM dynamics on pressure profiles.

\section{Radial profiles of the gas pressure}
\label{Profile}

\begin{figure}[h]
\begin{center}
\includegraphics[width=1\textwidth]{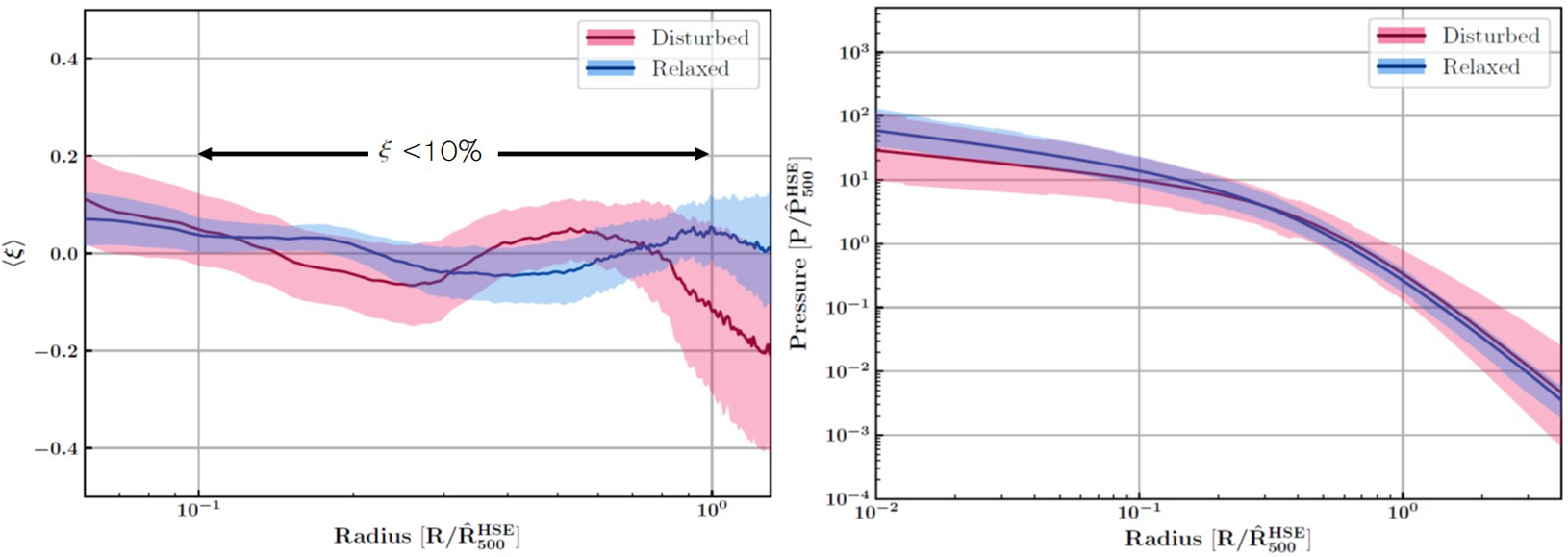}
\caption{{\it Left}: mean relative difference between the deprojected NIKA2/\textsl{Planck} and MUSIC pressure profiles along the normalized radius for the relaxed (blue) and the disturbed (red) populations. The shaded regions represent the 1$\sigma$ error on the mean. {\it Right}: mean normalized pressure profiles and associated 1$\sigma$ scatter obtained from the profile of relaxed (blue) and disturbed (red) clusters \cite{Florian19}.}
\label{fig-3}       
\end{center}
\end{figure}

The simulated tSZ maps are jointly analyzed with the NIKA2 tSZ analysis pipeline based on a MCMC procedure to recover deprojected gNFW radial pressure profiles. The procedure has already been  applied to PSZ2 G144.83+25.11, the first cluster of the NIKA2 SZLP \cite{Florian18}.
The impact of ICM disturbances is evident by studying the relative difference between the true pressure profiles derived from MUSIC-2 data and the recovered deprojected pressure profiles, $\xi$. See the mean difference profile with $\pm$1$\sigma$ error in the left panel in Fig.~\ref{fig-3}. This difference is always lower than 10\% in the radial region where the tSZ signal is not filtered significantly by NIKA2, independently on the dynamical state of the clusters. In fact, deviations at different positions are averagely compensated in the sub-sample of the disturbed clusters.
By analyzing the distributions of the normalized pressure profiles of the two populations separately, it is possible to estimate the mean pressure profile associated to relaxed and disturbed clusters, shown in the right panel of Fig.~\ref{fig-3}.

\begin{figure*}[h]
\centering
\includegraphics[scale=0.24]{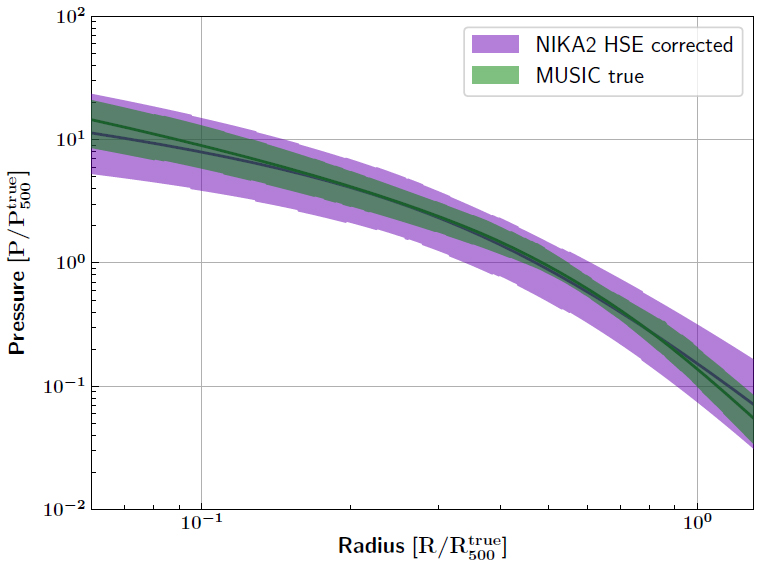}
\caption{Mean normalized pressure profiles and associated 1$\sigma$ for the whole {\sl twin sample}. The green curve represents the true pressure profile from MUSIC data. The purple curve gives the NIKA2/\textsl{Planck} deprojected profiles normalized by the integrated quantities corrected for the mean hydrostatic bias \cite{Florian19}.}
\label{fig-4}
\end{figure*}

As expected, the disturbed clusters show a larger scatter, mainly on the outskirts where the ICM disturbances, which are not spherically symmetric, do not accommodate the external slopes of the individual pressure profiles. 
Interestingly, the scatter of the mean profile of the disturbed clusters at $R_{500}$ is 65\% greater than the one observed for the relaxed clusters. 
Therefore, the intrinsic scatter associated with the distribution of pressure profiles is affected by the fraction of disturbed clusters, which also varies with the redshift.
Moreover, we compared the mean pressure profile of the full sample after a normalization by
corrected integrated quantities, $i.e.$ $R_{500}^{corr}$ and $P_{500}^{corr}$ taking into account an average bias in the estimated hydrostatic mass, with the true mean normalized pressure profile computed from the MUSIC-2 data. The profiles are consistent but the intrinsic scatter associated with the NIKA2/\textsl{Planck} deprojected profiles is on average twice as large as the MUSIC profiles, see Fig.~\ref{fig-4}. Most of this difference is explained by the simple spherical model for the thermal pressure used in our deprojection procedure that could fail in recovering disturbed cluster properties. It is worth to stress that even relaxed clusters could not satisfy the spherical model. The application of a triaxial deprojection methodology ($e.g.$ \cite{Sereno12}), combining NIKA2 SZ and XMM-$Newton$ X-ray data, would be an improved strategy to mitigate the impact of cluster morphology.

\section{Conclusions }
\label{Conclu}
A {\sl twin sample} of the current catalog of clusters selected for the NIKA2 tSZ Large Program has been extracted from the MUSIC hydrodynamical simulation. The synthetic clusters, spanning the same redshift and mass ranges of the SZLP, enable full access to all the ICM and dark matter data plus some additional information to segregate the objects by dynamical state. 
Realistic NIKA2 and \textsl{Planck} tSZ maps have been produced for each cluster. The NIKA2 tSZ pipeline ran on the mock observations to recover the pressure profile of each selected cluster, under the hypothesis of hydrostatic equilibrium. Mean normalised deprojected radial pressure profiles are derived for two sub-sample of clusters: relaxed and disturbed objects.
The capabilities of NIKA2, as a high angular resolution camera at millimetre wavelengths, allow us to detect the presence of ICM disturbances even at high redshifts up to $R_{500}$ highlighting the impact they have on both the shape and the scatter of the mean pressure profile, mainly in the case of disturbed clusters. We notice an increase of the intrinsic scatter of the pressure profile distribution at $R_{500}$ up to 65\% in the case of the disturbed population, which obviously depends on the morphological properties of the selected disturbed clusters. 

\section{Acknowledgements}
\label{Aknow}
This work is partially supported by the French National Research Agency under the contracts ANR-15-CE31-0017 and in the framework of the "Investissements d’avenir” program (ANR-15-IDEX- 02) and by funding from Sapienza Universit\'a di Roma - Progetti di Ricerca Medi 2017. We acknowledge fundings from the ENIGMASS French LabEx (F. R.). The MUSIC simulations were produced with the Marenostrum supercomputer at the Barcelona Supercomputing Centre thanks to computing time awarded by Red Espanola de Supercomputaci\'on. G.Y. acknowledges financial support by the MINECO/FEDER in Spain.

%
%

\end{document}